\documentstyle[sprocl]{article}

\input{psfig}
\def\be{\begin{equation}}
\def\ee{\end{equation}}
\def\bea{\begin{eqnarray}}
\def\eea{\end{eqnarray}}
\def\beq{\begin{equation}}
\def\eeq{\end{equation}}
\def\beqn{\begin{eqnarray}}
\def\eeqn{\end{eqnarray}}
\def\pl#1#2#3{{\it Phys. Lett. }{\bf #1}(19#2)#3}
\def\zp#1#2#3{{\it Z. Phys. }{\bf #1}(19#2)#3}
\def\prl#1#2#3{{\it Phys. Rev. Lett. }{\bf #1}(19#2)#3}

\def\pr#1#2#3{{\it Phys. Rev. }{\bf #1}(19#2)#3}
\def\np#1#2#3{{\it Nucl. Phys. }{\bf #1}(19#2)#3}

\newcommand\sss{\scriptscriptstyle}

\newcommand\as{\alpha_{\sss S}}

\def \as   {\mbox{$\alpha_s$}}

\def \bbbar {\mbox{$b \bar b$}}         
         
\def \pt   {\mbox{$p_{\scriptscriptstyle T}$}}                
\def \et   {\mbox{$E_{\scriptscriptstyle T}$}}

\def \mur  {\mbox{$\mu_{\rm \scriptscriptstyle{R}}$}}                
\def \muf  {\mbox{$\mu_{\rm \scriptscriptstyle{F}}$}}               
\def \muo  {\mbox{$\mu_0$}}                      
\def \mut  {\mbox{$\sqrt{p_{\sss T}^2+m_b^2}$}}
\def \to   {\mbox{$\rightarrow$}}

\begin{document}

{\flushright{
        \begin{minipage}{4cm}
        ETH-TH/96-23 \hfill \\
        hep-ph/9607333\hfill \\
        \end{minipage}        }

}

\title{BOTTOM PRODUCTION IN HADRONIC COLLISIONS~\footnote{To appear
in the proceedings of the {\it XI Topical Workshop on $p\bar{p}$ Collider
Physics}, Abano Terme, Italy, May 26$^{th}$ - June 1$^{st}$, 1996.}}

\author{STEFANO FRIXIONE}

\address{Theoretical Physics, ETH, Zurich, Switzerland}

\maketitle\abstracts{I review the status of the comparison 
between theoretical predictions and experimental results for 
bottom production in hadronic collisions, and discuss the 
possible sources of the discrepancies found. The study of
jets containing bottom quarks is proposed as a promising 
tool to investigate the $b$ production mechanism. I present
next-to-leading order QCD predictions for this process,
and compare them with data.
}

\section{Open bottom at fixed-target and collider experiments}

Bottom production constitutes a challenging testing ground
for perturbative QCD. The quark mass, which sets the scale of the hard
process, is such that $\as\simeq 0.2$. Therefore, the bottom rates 
cannot be predicted with full reliability at NLO in QCD, the radiative
corrections being of the same size of the leading-order contribution.
On the other hand, non-perturbative phenomena are expected to
play a less important r\^{o}le than in the case, for example, of charm.

Fixed-target experiments have in general too a low energy to
perform a statistically significant study of bottom production.
Most of the available data have been obtained in $\pi N$ collisions, 
and allow for a measurement of the total cross section~\cite{b_piN,bE653}. 
A measurement of the total cross section has 
also been performed in $pN$ collisions~\cite{bE605}.
Unfortunately, due to the limited coverage of the 
detectors, the results are somewhat model-dependent.
Taking into account the large theoretical uncertainties, the data
and the theory are in reasonable agreement. There is no value
of the bottom mass which allows to describe all the experimental
results; at variance with the measurements at colliders, some
results favour large values of the bottom mass.
The E653 collaboration~\cite{bE653} also presented a measurement
for single-inclusive and double differential distributions,
which turn out to be consistent with QCD predictions.

Bottom quarks are copiously produced at colliders. Although
the rejection of the background in the low-$p_{\sss T}$ region,
where most of the $b$'s are produced, is difficult, a large
set of data is available for distributions measured in the
central region in rapidity.

\begin{figure}
\centerline{\psfig{figure=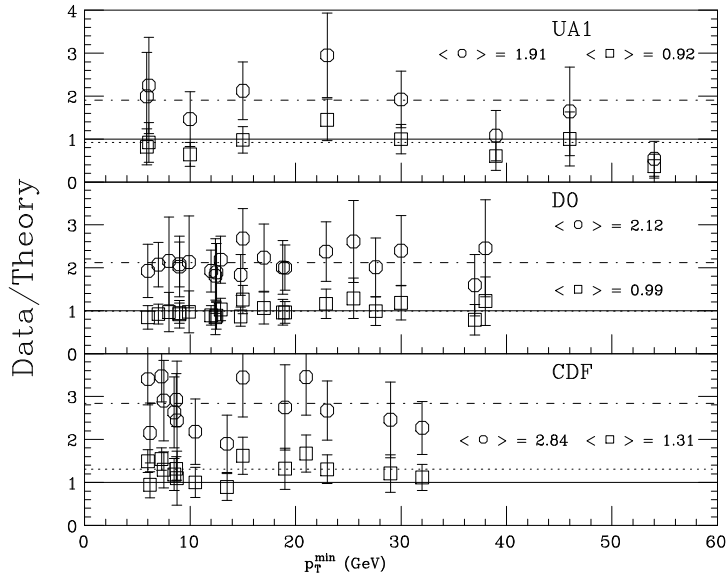,height=3.0in}}
\caption{Comparison between theory and data for bottom production
in $p\bar{p}$ collisions. See the text for details.
\label{fth_vs_exp}}
\end{figure}

In figure~\ref{fth_vs_exp} I present the comparison between the
NLO QCD predictions~\cite{HVQ_NLO} and the experimental 
results~\cite{b_at_coll} for the bottom cross section 
$\sigma\left(p_{\sss T}>p_{\sss T}^{min}\right)$ 
at $p\bar{p}$ colliders, as a function of $p_{\sss T}^{min}$.
The blobs have been obtained by dividing the data by the central
theoretical curve (the default values of the parameters entering
the calculations are: $m_b=4.75$~GeV, \mbox{$\mur=\muf=\muo$},
\mbox{$\Lambda_{5}^{\overline{\sss MS}}=152$~MeV}, where $\muo=\mut$
is the transverse mass of the bottom quark). The boxes are on the
other hand obtained by considering quite an extreme choice of the
parameters, namely $m_b=4.5$~GeV, $\mur=\muf=\muo/2$,
$\Lambda_{5}^{\overline{\sss MS}}=300$~MeV. This choice gives a result
which can be considered as the upper limit of the theoretical predictions.
I used the MRSA$^\prime$~\cite{MRSAP} set for the partonic densities; 
the dependence
of the result upon the choice of the densities is small, since they are 
known with a good accuracy in the $x$ range probed in $b$ production at 
Sp$\bar{\rm p}$S and Tevatron. A result quite close to the
upper limit can also be obtained by choosing $m_b=4.5$~GeV, 
$\mur=\muf=\muo/4$, $\Lambda_{5}^{\overline{\sss MS}}=152$~MeV.
Further details can be found elsewhere~\cite{fmnr}.
The average value of the {\it data/theory} points (represented in the
figure by the dotted and dot-dashed lines) has been calculated
weighting these points with the inverse of their relative error.

It is apparent from the figure that the results at $\sqrt{S}=630$~GeV
and at $\sqrt{S}=1800$~GeV are consistent with each other.
The CDF results are about 30\% higher in normalization than the D0 ones,
while the shape of the theoretical curves is well reproduced
by all the three collaborations. The comparison with the theory is
quite satisfactory if one is willing to accept an extreme choice of
the parameters: collider measurements favour small values of the
bottom mass and of factorization/renormalization scales, and $\as$ values 
compatible with LEP measurements. On the other hand, the data are higher 
than the default theory prediction by a factor of 2 or more.

The fixed-order perturbative QCD calculations I used above to compare
with the experimental data may become unreliable in certain kinematical
regions, due to the appearance of potentially large logarithms which
spoil the convergence of the perturbative expansion. In this case,
a resummation to all orders of these large logarithms has to be performed.

When the available center-of-mass energy gets large, the effective
expansion parameter of the perturbative series becomes
\mbox{$\as\log(S/m_b^2)$}. The problem of resumming these terms
({\it small}-$x$ {\it effects}) has been tackled by several
authors~\cite{res_smallx}. In the Tevatron energy regime, it was shown
that the total cross section can increase by a factor of 30\% at most
with respect to the NLO prediction.
Furthermore, the resummation should have a negligible effect
in the tail of the $p_{\sss T}$ distribution, where the effective
scale is not the quark mass, but the transverse mass, and the ratio 
$S/\mu_0^2$ is not that large.

The transverse momentum distribution is in principle more affected 
by the presence of~\mbox{$\log(p_{\sss T}/m_b)$} terms. These logarithms
can be resummed by observing that, at high $p_{\sss T}$, the bottom
mass is negligible, and by using perturbative fragmentation 
functions~\cite{res_highpt}. It turns out that the resummation only 
slightly changes the shape of the fixed-order prediction, but improves 
the perturbative stability of the result.

Finally, multiple soft gluon emission makes the perturbative expansion
unreliable close to the threshold or to the borders of
the phase space, like for example the regions 
$p_{\sss T}^{b\bar{b}}\simeq 0$ and $\Delta\phi^{b\bar{b}}\simeq \pi$.
A lot of theoretical work has been performed in this
field~\cite{res_softgl}; at currently probed energy, these effects
are not affecting the total rate or single inclusive distributions,
while they may be relevant when investigating more exclusive quantities,
like the correlations between the quark and the antiquark.

\begin{figure}
\centerline{\psfig{figure=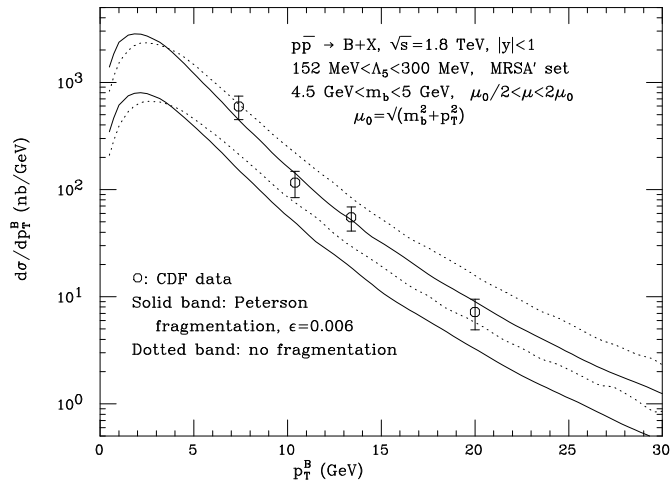,height=2.5in}}
\caption{Transverse momentum spectrum of the $B$ mesons: CDF
data versus theoretical predictions.
\label{fcdf_band}}
\end{figure}
One is therefore led to conclude that the resummation of large
logarithms cannot improve the comparison between theory and data 
for the $p_{\sss T}$ spectrum at the Sp$\bar{\rm p}$S and Tevatron
colliders. It has to be observed that experimental results for
bottom quarks depend on the assumptions made for the hadronization
process, since only $B$ hadrons are experimentally accessible.
On the other hand, to compare with data on $B$ mesons, 
the QCD prediction for bare quarks has
to be convoluted with a fragmentation function. Usually, the
fragmentation function is determined by fitting $e^+e^-$ data.
In this way, it is conceivable that the fraction of $B$ mesons
coming from the splitting $g\to b\bar{b}$ is underestimated in hadron 
collisions, since this mechanism is much more important in this case
than in the case of $e^+e^-$ annihilations. I will show later that
indeed the gluon splitting is a key feature for bottom production
at Tevatron. In figure~\ref{fcdf_band} the CDF data~\cite{CDF_Bmes}
on the $p_{\sss T}$ spectrum of $B$ mesons are compared with the QCD 
predictions obtained with and without the Peterson fragmentation.
The experimental measurements are close to the upper limit 
of the theoretical curve with fragmentation. On the other hand,
they stay inside the band obtained without fragmentation, 
displaying a slightly softer behaviour. Notice that, although the
fragmentation cannot affect the total rate, it however does
affect quantities like $\sigma\left(p_{\sss T}>p_{\sss T}^{min}\right)$ 
for moderate and large $p_{\sss T}^{min}$ values, since the degradation
of the momentum is sizeable in the tail of the
transverse momentum spectrum. For this reason, it would be
useful to have data on the transverse momentum of the $B$ mesons 
for $p_{\sss T}$ values larger than those displayed in 
figure~\ref{fcdf_band}. This would help in clarifying the issue
whether in hadronic collisions the experimental results favour
a fragmentation function more peaked towards the region $x\simeq 1$
than suggested by the Peterson parameterization.

The very same considerations enter into play when the comparison
between theory and experiments is made for more exclusive
quantities. As mentioned before, in this case the importance
of soft gluons effect has to be taken into account. The CDF
collaboration recently presented~\cite{CDF_corr} a study on
$b\bar{b}$ correlations at the Tevatron, and the results appear
to be at variance with QCD. Since correlations provide
us with the most complete information on the production mechanism,
further studies should be devoted to this topic.

\section{Heavy-quark jets}

An interesting way of understanding the production
mechanism of heavy flavours is to consider the cross section of
jets which contain a heavy quark~\cite{FM} (briefly:
heavy-quark jets).  The main difference between the
study of a heavy quark and a heavy-quark jet is that in the former case one is
interested in the momentum of the quark itself, regardless of the properties of
the event in which the quark in embedded, while in the latter case one is
interested in the properties of a jet containing one or more heavy quarks,
regardless of the momentum fraction of the jet carried by the quark. A priori
it is expected that variables such as the \et\ distribution 
of a heavy-quark jet should be described by a finite-order QCD calculation 
more precisely than the \pt\ distribution of open quarks,
since the jet \et\ does not depend on whether the energy is carried all 
by the quark or is shared among the quark and collinear gluons, and therefore
large collinear logarithms $\log(\pt/m_b)$ do not appear in the cross section.
The experimental measurement of the \et\ distribution of 
heavy-quark jets does not depend on the knowledge 
of the heavy-quark fragmentation functions, contrary to the case 
of the \pt\ distribution of open heavy quarks. 
Experimental systematics, such as
the knowledge of decay branching ratios for heavy hadrons or            
of their decay spectra, are also largely reduced.

The calculation of the heavy-quark jet rate is very similar to
the one of the generic jet cross section.
Two important differences have nevertheless to be stressed:
by its very definition, a heavy-quark jet is not flavour-blind; one 
has to look for those jets containing a heavy flavour. Furthermore, the
mass of the heavy flavour is acting as a cutoff against final
state collinear divergences. This in turn implies that the structure
of the singularities of the heavy-quark jet cross section is identical
to the one of the open-heavy-quark cross section (a proof of this statement,
and a detailed derivation of all the steps needed to build a NLO
heavy-quark jet cross section in perturbative QCD, can be found
elsewhere~\cite{FM}). The heavy-quark jet cross section at NLO
can therefore be written in the following way:
\beq
d\sigma=d\sigma^{(open)}+d\Delta\,,
\label{xsecsplit}
\eeq
where $d\sigma^{(open)}$ is the open-heavy-quark cross
section, and $d\Delta$ is implicitly defined in 
eq.~(\ref{xsecsplit}). The key feature of this equation is that
all the subtractions needed to get an infrared-safe result
are contained in the term $d\sigma^{(open)}$. By construction, 
at NLO in QCD a heavy-quark jet can coincide with the heavy quark 
itself, or it can contain a heavy quark and a light parton, or the
heavy quark-antiquark pair. The latter two possibilities are peculiar
of the heavy-quark jet cross section, and are not present in the
open-heavy-flavour one; formally, they are described by the
$\Delta$ term in eq.~(\ref{xsecsplit}), which contribution I will
call from now on as ``jet-like component'' of the cross section.

To present some results of interest for measurements
at the Tevatron~\cite{koehn},
I will consider jets produced within $|\eta|<1$, in order to 
simulate a realistic geometrical acceptance of the Tevatron detectors. 
The jets will be defined using the Snowmass convention~\cite{snowmass}, 
whereby particles are clustered in cones of radius $R$ in the 
pseudorapidity-azimuthal angle plane. The default parameters are the 
same as before, but now \mbox{$\muo=\sqrt{E_{\sss T}^2+m_b^2}$}, where
\et\ is the transverse energy of the $b$-jet (notice that \et\ is not
equal to \pt\, since the bottom is massive; the difference is however
almost negligible in the energy range interesting for current
phenomenological studies).

\begin{figure}
\centerline{\psfig{figure=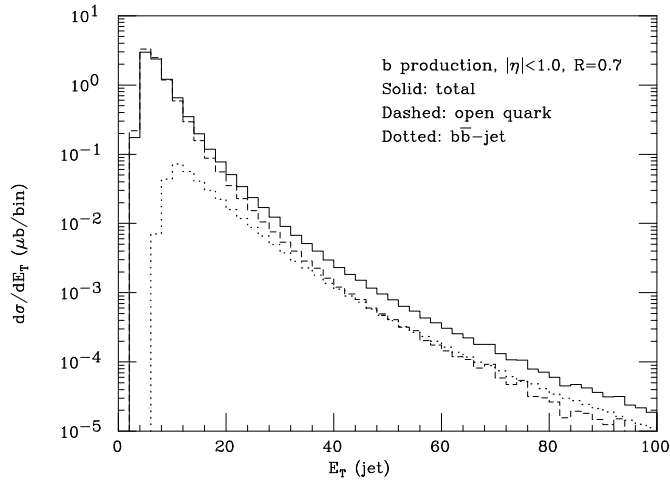,height=2.5in}}
\caption{$b$-jet inclusive \et\ distribution in $p\bar p$ 
collisions at 1.8 TeV, for 
$|\eta|<1$, $R=0.7$ and $\muf=\mur=\muo$ (solid line). 
For comparison, the open-quark inclusive \et\ distribution 
(dashed line) is also presented. The component of the jet-like
contribution due to jets containing both $b$ and $\bar b$ is 
represented by the dotted line.
\label{fbinc}}
\end{figure}
Figure~\ref{fbinc} shows the prediction for the \et\ distribution 
of $b$-jets at the Tevatron for $R=0.7$. For the
purpose of illustration, the open-quark component is separately presented. 
It is apparent that the jet-like component
becomes dominant as soon as \et\ becomes
larger than 50~GeV. It can be shown~\cite{FM} that this values 
actually depends significantly on the cone size, 
being equal to 25 and 100~GeV for $R=1$ and 0.4 respectively.
I also show the part of the jet-like component due to
jets that include the \bbbar\ pair (I will call these $\bbbar$-jets). 
The figure suggests that, for this \et\ range and 
with $R=0.7$, this is the dominant part of the jet-like component.
This is consistent with the expectation that, for large enough \et\  and
provided  that the majority of the final-state generic jets are composed of
primary gluons, heavy-quark jets are dominated by the process of gluon
splitting, with the jet formed by the heavy-quark pair. 
\begin{figure}
\centerline{\psfig{figure=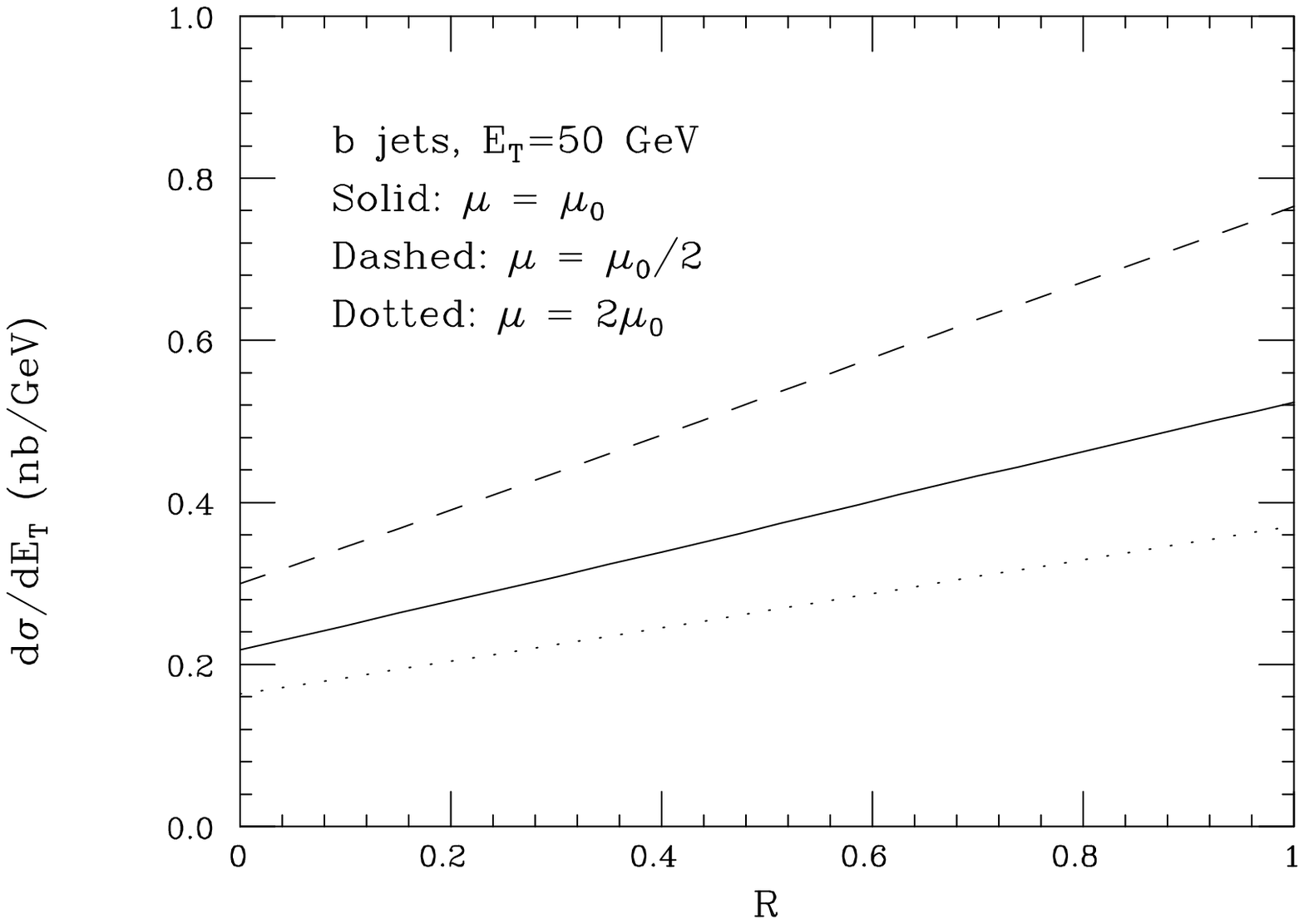,height=1.7in}
            \hspace{0.3cm}
            \psfig{figure=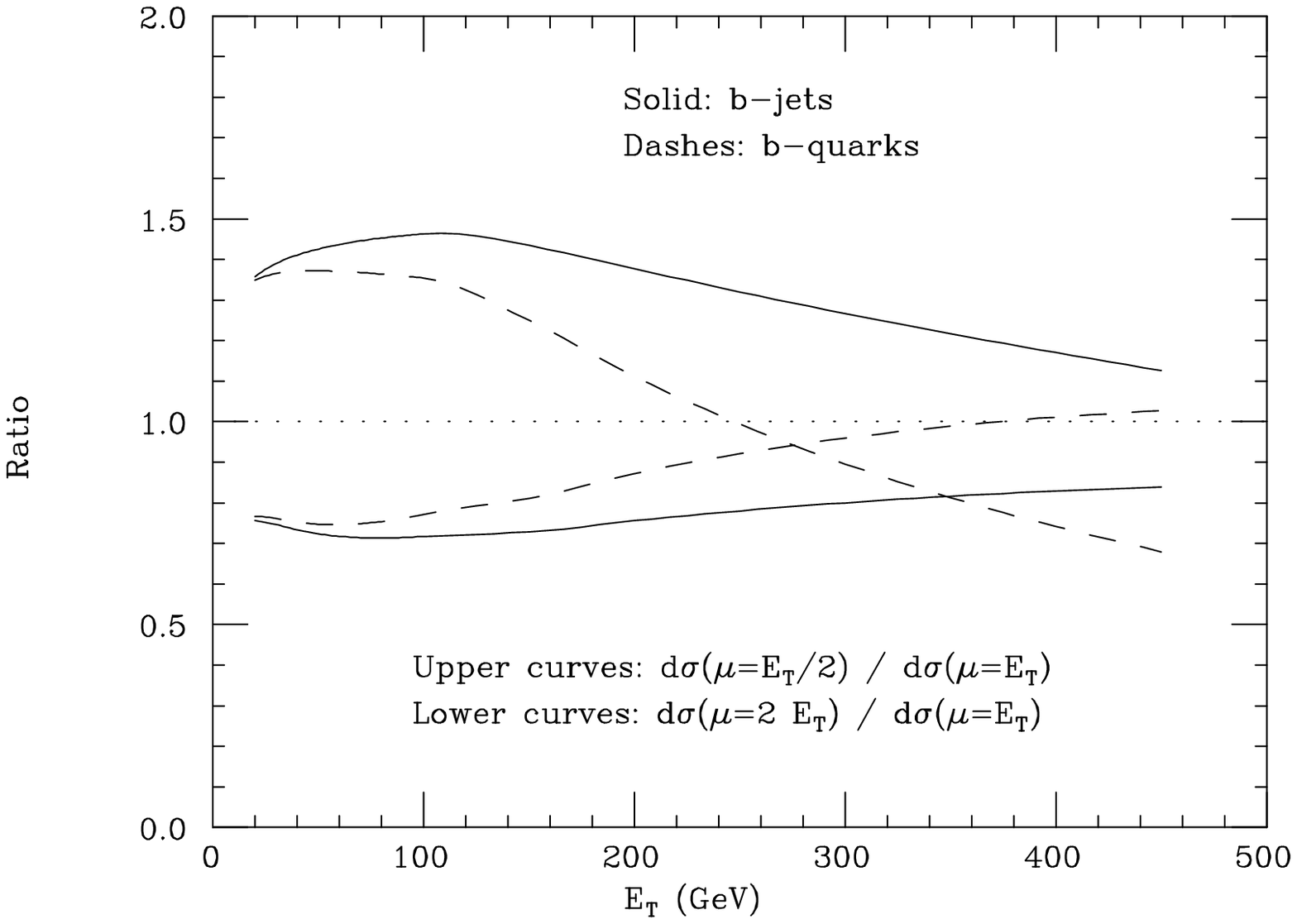,height=1.7in}}
\caption{Left: $b$-jet inclusive \et\ rate, as a function of the cone size 
$R$, at \et~=~50~GeV and for various scale choices ($\mur=\muf\equiv\mu$).
Right: Scale dependence of the $b$-jet \et\ distribution ($R=0.4$, 
solid lines) and of the open-quark inclusive \et\ distribution 
(dashed lines).
\label{fbscale}}
\end{figure}

The left side of figure~\ref{fbscale} presents the theoretical prediction
for the absolute heavy-quark jet rate at $\et=50$~GeV versus the
cone size, for different choices of the factorization/renormalization scale.
In this case, the cross section at $R=0$ is well defined, and it is equal 
to the open-heavy-quark cross section. This should be contrasted with the case 
of generic jets, in which the cross section at $R=0$ is not well defined, 
being negative at any fixed order in perturbation theory~\cite{nlojets}.
The right side of figure~\ref{fbscale} shows the scale dependence of the 
$b$-jet cross section (for $R=0.4$) as a function of \et, 
for values up to 450~GeV.

The strong scale dependence exhibited by the absolute rates at low and
moderate \et\ values is of the same size as the one present in the 
inclusive \pt\ distribution of open bottom quarks.
This scale dependence is usually attributed to the 
importance of the gluon splitting contribution. 
One expects therefore that in a regime in which the gluon
splitting contribution is suppressed by the dynamics the scale dependence
should be milder. I will show later that this suppression is indeed
taking place for high transverse energies. This explains why 
in the high-$\et$ region the scale dependence is indeed reduced to the
value of $20$\% when the scales are varied in the range $\muo/2<\mu<2\muo$, 
a result consistent with the limited scale dependence of the NLO
inclusive-jet cross sections~\cite{nlojets}.

\begin{figure}
\centerline{\psfig{figure=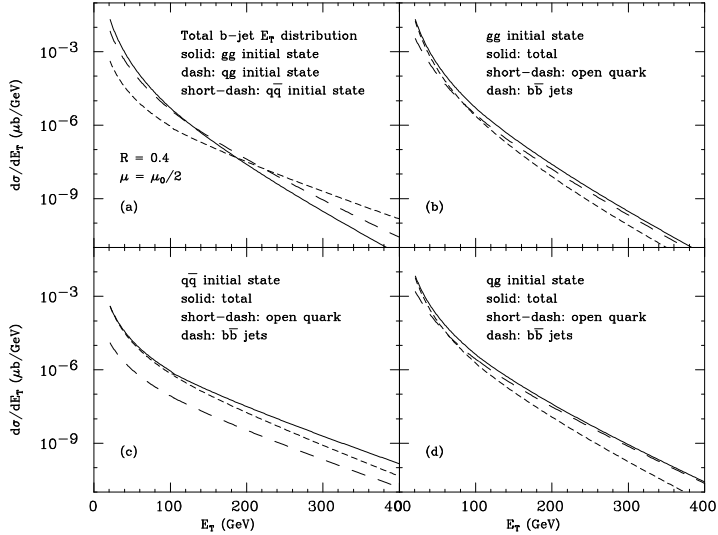,height=2.8in}}
\caption{Initial state composition of the $b$-jet production processes,
calculated for $\mur=\muf=\muo/2$ and $R=0.4$ (upper left).
Different components of the                             
production processes: $gg\;\to\; b$-jet (upper right), $q{\bar q} \;\to\;
b$-jet (lower left) and $qg\;\to\; b$-jet (lower right).
\label{fbhtot}}
\end{figure}

The high-energy behaviour of $b$-jet cross section is presented in 
figure~\ref{fbhtot}, for a given choice of scales and cone size.
In fig.~\ref{fbhtot}a the separate contribution to the $b$-jet 
cross section of the three possible initial states, $gg$, $q\bar q$ 
and $qg$, is displayed. Notice that the
$q\bar q$ contribution becomes dominant for $\et>250$~GeV.
Figures~\ref{fbhtot}b--d show, for each individual channel,  the separate
contribution of the open-quark and \bbbar-jet components. For \et\ large
enough, the dominant component of the $gg$ and $qg$ channels is given by the
\bbbar-jet contribution, because of the gluon-splitting dominance. In the case
of the $q\bar q$ channel, on the contrary, the \bbbar-jet term is always
suppressed, and most of the $b$-jets are composed of a single $b$
quark, often accompanied by a nearby gluon. 

\begin{figure}
\centerline{\psfig{figure=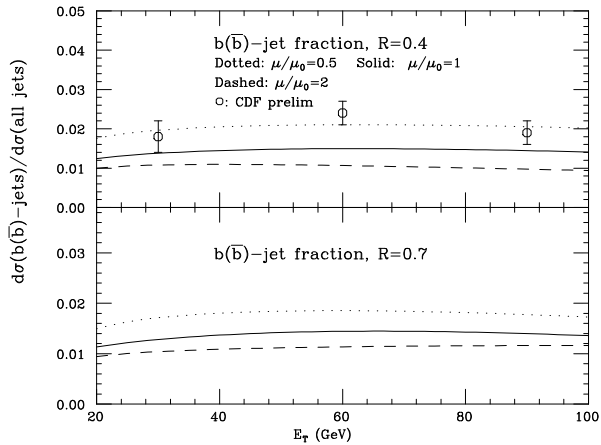,height=2.3in}}
\caption{Ratio of the $b(\bar b)$-jet to inclusive-jet \et\ 
distributions, for different choices of       
renormalization and factorization scales ($\mur=\muf\equiv\mu$), 
for $R=0.4$ (top) and
$R=0.7$ (bottom). The data points for $R=0.4$ represent preliminary results
from the CDF experiment,  for which only the statistical
uncertainty is shown.
\label{fbjetfrac}}
\end{figure}

Coming finally to the comparison with data, preliminary
results are available~\cite{koehn} for the fraction of heavy-quark jets 
relative to generic jets. I present in fig.~\ref{fbjetfrac} the ratio of 
the $b$-jet to inclusive-jet \et\ distributions. The 
inclusive-jet \et\ cross section has been calculated with the JETRAD
program~\cite{jetrad}. For consistency with CDF prescriptions~\cite{koehn},
the $b$-jets are defined here as jets containing either a $b$ or a $\bar b$
quark, jets containing both being counted only once. I will call these
$b(\bar b)$-jets. 
It is interesting to notice that there is a good agreement between 
the CDF data and the theoretical prediction obtained with 
$\mur=\muf=\muo/2$; this choice of scale is also supported
by inclusive-jet \et\ spectrum~\cite{jetdata,jetdataD0} data.
This is particularly significant since the choice of scale for the
heavy-quark jet cross section is not independent from the scale chosen to 
predict the open-heavy-quark one (see eq.~(\ref{xsecsplit})).
But, as I stressed before, to get a satisfactory description of
the data for the open-bottom \pt\ spectrum, a more extreme choice
of the parameters has to be done. Should this situation persist when
additional data on $b$-jets will become available, it would
indicate an inconsistency in describing two phenomena due to
the same underlying physics. The poor understanding of the 
fragmentation mechanism is very likely a source of this inconsistency.
In this sense, the study of $b$-jet production is a very promising
tool, since theoretical predictions are in this case independent
from a detailed knowledge of the final state long-distance physics.

\section*{Acknowledgments}
I would like to thank Michelangelo Mangano, Paolo Nason and Stefano
Passaggio for useful discussions. Financial support by the National Swiss
Foundation is also acknowledged.

\end{document}